# Evaluation of a College Freshman Diversity Research Program

**Sarah Garner**
University of Washington, Seattle, Washington 98195
**Michael J. Tremmel**
University of Washington, Seattle, Washington 98195
**Sarah J. Schmidt**
The Ohio State University, Columbus, Ohio 43210
**John P. Wisniewski**
University of Oklahoma, Norman, Oklahoma 73019
**Eric Agol**
University of Washington, Seattle, Washington 98195

Abstract:
Since 2005, the Pre-Major in Astronomy Program (Pre-MAP[1]) at the University of Washington (UW) Department of Astronomy has made a concentrated effort to recruit and retain underrepresented undergraduates in science, technology, engineering and mathematics (STEM). This paper evaluates Pre-MAP in the context of the larger UW student population using data compiled by the University's student database. We evaluate the Pre-MAP program in terms of our goals of recruiting a more diverse population than the University and in terms of a higher fraction of students successfully completing degrees. We find that Pre-MAP serves a higher percentage of underrepresented minorities and equal percentages of women compared to entering freshmen classes at UW. Additionally, Pre-MAP has a higher percentage of degree completion with higher average GPA's and similar time to completion when compared to UW as a whole and other STEM majors, particularly with students that place into lower-level math courses (such as basic algebra or pre-calculus).

## 1. INTRODUCTION
Since 2005, the Pre-Major in Astronomy Program (Pre-MAP) at the University Of Washington (UW) Department Of Astronomy has made a concentrated effort to recruit and retain underrepresented undergraduates in science, technology, engineering and mathematics (STEM). Previous work regarding the evaluation of the program has been focused on small-scale statistics and impacts on individuals, such as measuring student's confidence levels of programming before and after Pre-MAP (Haggard, Garner & Cowan 2008). This paper shares the results of the first large-scale quantitative comparison using data for the entire UW student population. We will compare academic performance, recruitment, retention, and degree completion between Pre-MAP students, Physics/Astronomy majors, STEM majors, and the University at-large. The goal of this evaluation is to make specific programmatic recommendations for improvement and growth.

The guiding questions for this evaluation are as follows:
   1) Does Pre-MAP recruit a diverse population of students compared to the University?

---
[1] Pre-MAP is supported by NSF Career grant AST-0645416 and the Department of Astronomy at the University of Washington. This evaluation was supported in part by NSF AST-100931.

2) How does math placement of Pre-MAP students impact STEM retention? How do Pre-MAP students perform in STEM courses compared to other non-Pre-MAP students?
3) How does academic performance impact time-to-degree for Pre-MAP students in comparison with the University?
4) Are Pre-MAP students choosing and completing STEM majors more often than other UW students?

**1.1 Pre-MAP Structure**

Pre-MAP[2] is a research seminar offered each autumn quarter. It is intended primarily for incoming freshmen and has occasionally included sophomores and transfer students. We recruit students using a variety of methods, including an email advertisement and contact with students during incoming academic advising sessions.

The first half of the quarter is an intensive computing seminar to increase the students' familiarity with programming skills useful for science research. During the second half of the quarter, the students focus entirely on research projects. Each group of two or three students works on a single project under the guidance of a faculty member, postdoc or graduate student. At the end of the quarter, all Pre-MAP students present their research projects to the Astronomy Department.

The classroom portion of Pre-MAP ends after one quarter, but all students are encouraged to continue research projects after the completion of the seminar if they have an interest in doing so. Many Pre-MAP students also continue to receive informal mentoring from the Pre-MAP staff and/or their research mentors from the seminar. As long as they are enrolled in UW, Pre-MAP students are encouraged to attend events and field trips organized as part of the extended program. Some students continue with research, informal mentoring, and extended program events until they graduate, while other students choose to move on after the course is over.

The first Pre-MAP seminar was offered in Fall 2005, and the program has served 73 students as of the Fall 2011 seminar, averaging cohorts of 10 per year. While the program is small in size compared to the UW undergraduate student population (roughly 30,000), each year's 10 student Pre-MAP cohort is comparable in size to the 20 astronomy majors and 50 physics majors per year.

**1.2 Program Goals**

Pre-MAP was designed to increase the number of under-represented students who chose to major in Astronomy or STEM fields. As we cannot directly test that goal due to relatively small numbers and the lack of a comparison sample, here we aim to test the recruitment and academic success of Pre-MAP students compared to other students at UW.

Our recruitment goal is simply to recruit a more diverse population of students than UW as a whole. Recruiting a diverse cohort ensures that any benefits of the program are directed at under-represented students.

---

[2] Additional information on the structure, programmatic details and administration of Pre-MAP can be found online: http://www.astro.washington.edu/users/premap/diy/premap.html (Rosenfeld et al. 2010).

The academic goals of Pre-MAP are multi-faceted. The most basic goal is to increase the number of underrepresented minorities (URM) graduating 1) with any degree; 2) with a degree in a STEM; or 3) with a degree in Astronomy and Physics. Our secondary goals are to support Pre-MAP students so that when compared to the average UW student their academic performance (GPA)is equal to or greater than and time to degree completion are equal to or less than. This would indicate that the Pre-MAP focus on research does not interfere with student academics.

## 1.3 Data

The University of Washington Student Database (UW SDB) is the central system used to track student records. The data includes several linked tables that track admissions, academic records, major declaration, and graduation data. As part of these tables, the UW SDB also tracks demographic data. The tables and fields containing the data are not all clearly documented and some have changed over the years relevant to our paper. Data from UW SDB was queried using MS Access in July 2012 and includes student records from Autumn 2005 through the spring of 2012.

We used four different tables to obtain the relevant data: degrees conferred, student demographics, transcripts and major declaration. In each table, individual students are identified by a system key, versus official student number, which allows us to link together data from each table while maintaining student anonymity. Pre-MAP students were identified by obtaining a list of their system keys. We then used custom IDL code to convert the data to fits tables, link together relevant entries, and create tables and figures used here.

## 2. RECRUITMENT

To increase the number of URM students who chose to major in STEM fields, Pre-MAP must first recruit a diverse group of students. We used the demographic ethnicity field (a self-identified field a student provides during admissions) in UW SDB to select students that match the National Science Foundation's definition of an underrepresented minority[3] within both the Pre-MAP and the UW population as a whole. In the years 2005-2011, Pre-MAP recruited a higher percentage of URMs and an equal percentage of women compared to the incoming UW class (*Table 1*). This indicates that our targeted recruiting techniques, as a whole, have been successful.

*Table 1*

|  | 2005 – 2012 | | 2005 – 2008 | | 2009 – 2012 | |
|---|---|---|---|---|---|---|
|  | Pre-MAP | Total UW | Pre-MAP | Total UW | Pre-MAP | Total UW |
| All Students | 73 | 102988 | 45 | 57151 | 28 | 54878 |
| All Ethnic Minorities | 25% | 10% | 29% | 9% | 18% | 9% |
| African American | 10% | 3% | 13% | 3% | 4% | 3% |
| Hispanic | 10% | 5% | 9% | 1% | 11% | 5% |
| American Indian | 4% | 1% | 4% | 1% | 4% | 1% |

---

[3] National Science Foundation "Women, Minorities, and Persons with Disabilities in Science and Engineering" http://www.nsf.gov/statistics/wmpd/, October 2012

| | | | | | | |
|---|---|---|---|---|---|---|
| Hawaiian/Pacific Islander | 1% | 1% | 2% | 0% | 0% | 1% |
| Women | 56% | 53% | 56% | 53% | 57% | 54% |

*Table 1: Pre-MAP and UW entering student demographics for the entering classes of 2005 - 2012. To test the change in recruitment over time, we include demographics 2005-2008 and 2009-2011. Division of years is arbitrary.*

Between 2005 and 2012, the Pre-MAP staff changed our recruiting techniques. Initial recruitment included planetarium shows to various middle and high school programs (e.g. Upward Bound) and attendance at many UW events designed for incoming students. These events were labor-intensive and did not produce sufficient return compared to the staff hours invested. During later years, we narrowed our recruiting focus and relied more on email advertisements and recruitment at incoming student orientation advising sessions.

To test the change in our recruitment over time, we compared the recruitment of URM and women in the years 2005-2008 and 2009-2012. We find that over time, the gap is closing between the percent of ethnic minorities in Pre-MAP recruitment and the percent in UW's freshman enrollment. There was no change in the percentage of women enrolled in Pre-MAP. The slight reduction in the percent of ethnic minorities in Pre-MAP could be due to our change in recruiting techniques. Additionally, the reduction of the percent of ethnic minorities in Pre-MAP could be due to the establishment of the Louis Stokes Alliance for Minority Participation (LSAMP)[4] program at UW in 2009. LSAMP serves as a major resource for URMs in STEM by connecting students to more programs. LSAMP students may be aware of more opportunities than students in previous years resulting in them selecting other diversity programs more aligned with their STEM interest versus an Astronomy-specific program.

### 3. OVERALL ACADEMIC PERFORMANCE

There are many different components to evaluating overall academic performance. Our initial exploration of academic performance focuses on grades and time to degree completion.

First, we will examine overall GPA at the time of graduation. We adopted the GPA field and the graduation major fields recorded in the degrees conferred data table. We identified STEM majors using a list produced by UW. *Figure 1* shows the distribution of GPA's for Pre-MAP, STEM, all UW, and Physics/Astronomy students. The distributions of grades are very similar across all majors, with only the ASTR majors having a slightly higher mean and smaller standard deviation than the rest of the students. The Pre-MAP GPA's are not significantly different than the GPA's of all UW, STEM, Physics, or Astronomy majors.

---

[4] http://depts.washington.edu/lsamp/

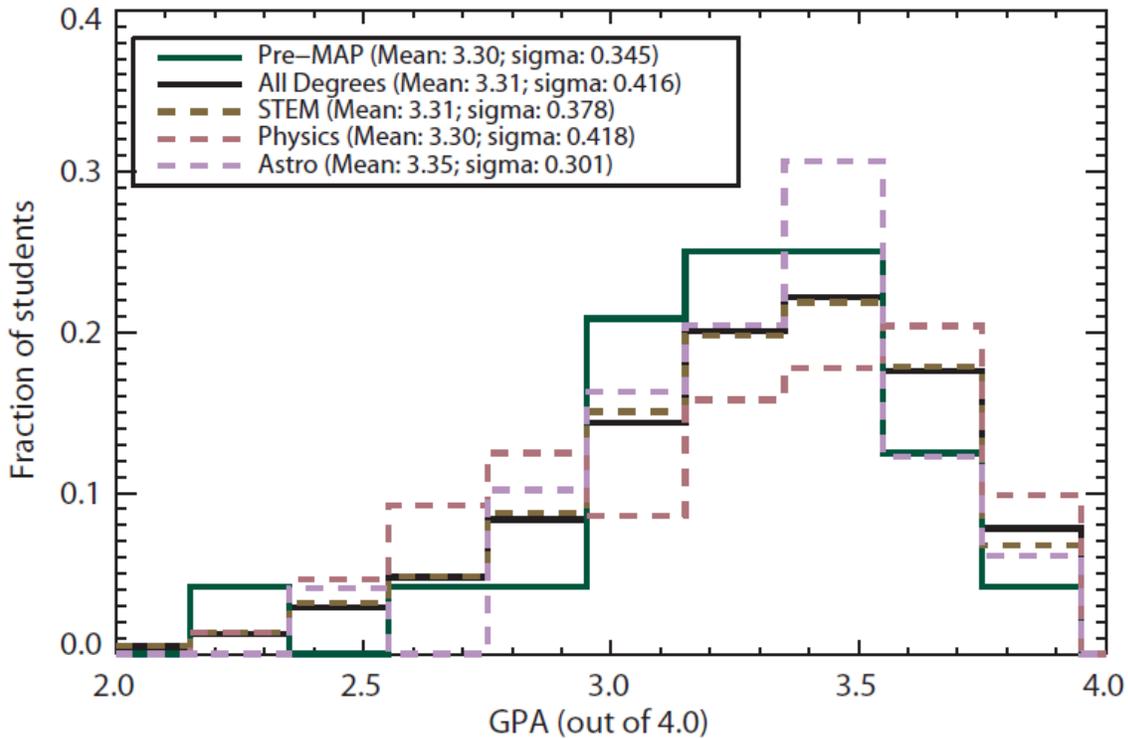

*Figure 1:* Cumulative GPA upon graduation for Pre-MAP, all UW, STEM, Physics, and Astronomy graduates. The means and standard deviations for each group are given in the legend. The similar distributions indicate that Pre-MAP students perform similarly overall compared with other relevant student populations.

While the overall GPA's of Pre-MAP students are the same as those in physics and astronomy majors, performance in coursework for a specific major is another important measure of academic success. *Figure 2* shows the distribution of upper level Physics, Astronomy, and Math GPA's for students who have taken six or more classes in upper level (>223) undergraduate Physics classes. The GPA's were calculated using the course grades recorded in the course data table. Each course is clearly identified by discipline and course number. We use the criterion of more than six or more upper level Physics courses to define the population of students who intend to major in physics. The minimum course numbers were selected to include major-specific courses and exclude introductory material. In addition, we define a subset within the "physics student" population to account specifically for those likely to pursue an astronomy degree. We do this by requiring the student having taken four or more upper level (>200) astronomy courses in addition to the upper level physics courses.

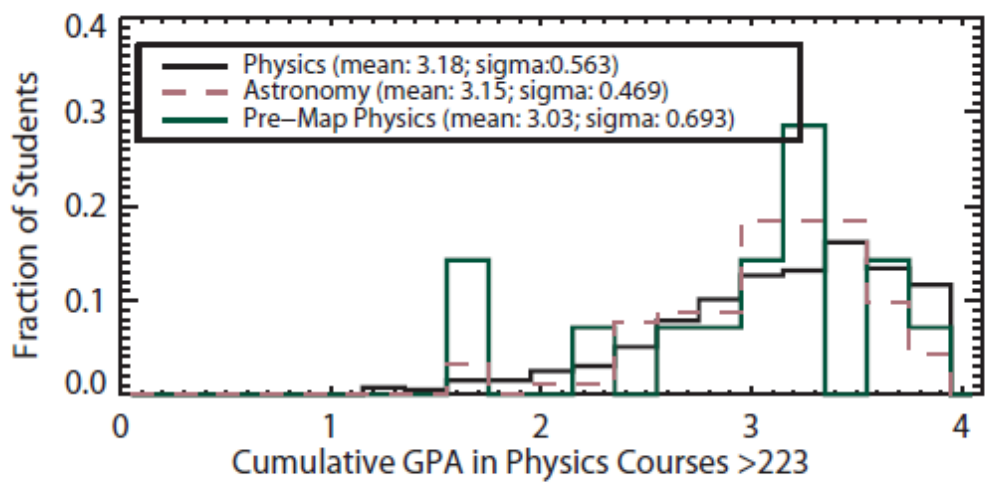

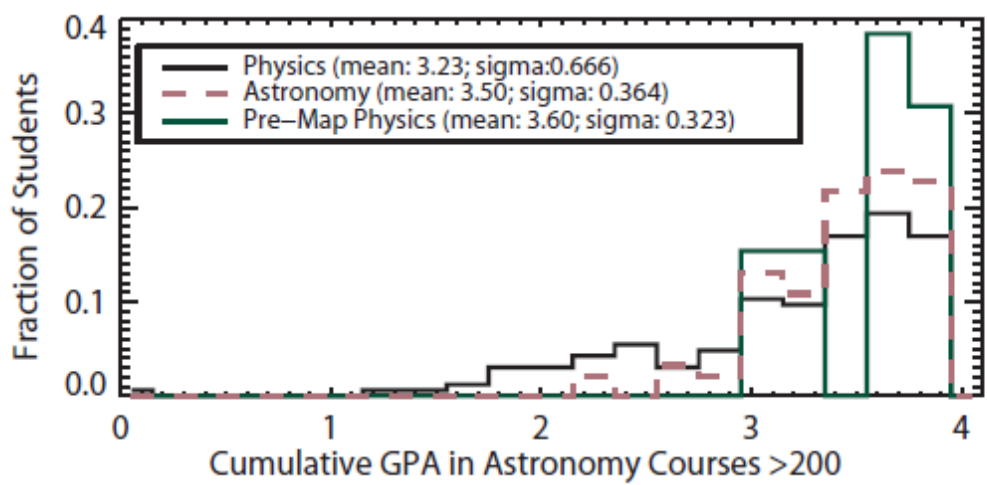

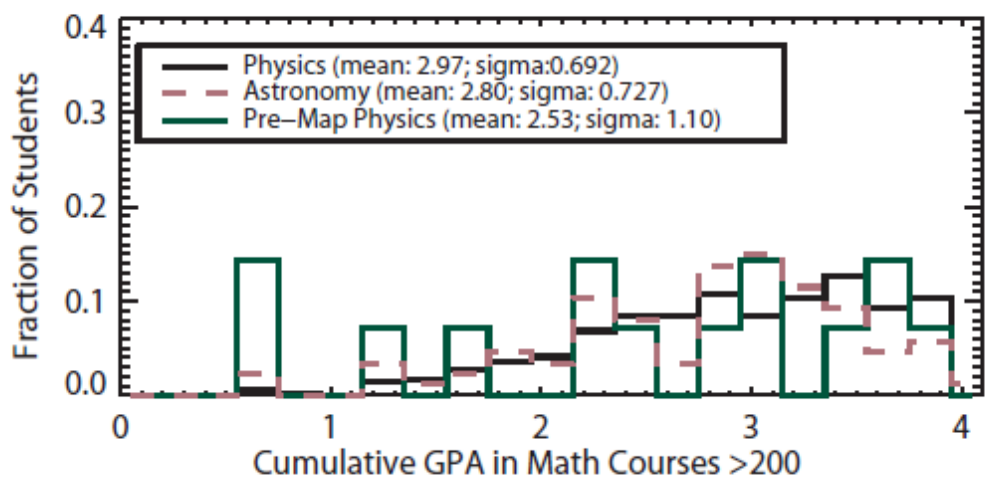

*Figure 2:* Cumulative GPAs in Physics, Astronomy, and Math are shown for students who have taken six or more upper level (>223) undergraduate physics courses. Pre-MAP students who meet the same criteria are shown (red line). Pre-MAP students underperform in physics courses compared with their peers, but they perform similarly to their peers in math courses (with a few notable outliers at low gpa) and perform better in Astronomy courses. Astronomy students are shown in the dashed line. They are defined to have taken six or more upper level physics courses as well as four or more upper level (>200) undergraduate Astronomy courses. We find that in all areas Pre-MAP students perform more similarly to the Astronomy students compared to the overall Physics student population.

Pre-MAP students are receiving lower grades in upper level physics classes, as indicated by a mean GPA of 2.94 for Pre-MAP students compared to 3.28 for all students. Pre-MAP students perform similarly in math courses (GPA of 3.09 compared to all GPA of 3.13) and excel in astronomy courses (GPA of 3.58 compared with 3.24). Pre-MAP students perform similarly to Astronomy students in all areas, with a couple notable outliers with low math gpa.

We speculate that Pre-MAP students are doing well in Astronomy compared to all students due to the cohort building component of Pre-MAP and early exposure to Astronomy in their first two years. Since they take Pre-MAP along with an introductory astronomy course and participate in research all before declaring an astronomy major, they already have lots of exposure to astronomy before taking the upper-division courses. This is in comparison with non-Pre-MAP students whose first two years are often filled with physics, math and general education requirements; many times their first exposure to academic astronomy is in upper-division classes.

**5. EXAMINING MATH PLACEMENT**
A student's math background and subsequent performance in math classes can be an important indicator of their future success in STEM. In this section we will study how the Pre-MAP student population compares to the overall UW population in math preparation prior to college, inferred by a student's initial math placement, and performance in math classes at UW. We will also examine the connection between math performance and the retention rate of students in STEM fields.

Incoming UW students are placed in a math course either through placement testing, AP credit, or transfer courses. To fulfill the math requirement, there are a variety of different math courses a student can take. Students who express interest in STEM are directed to the pre-calculus or calculus series. The first course they enroll in can be used as a proxy for their math placement, indicating high school preparation. We selected their first math course using the student course table. The numbers and descriptions of the courses, in addition to the fraction of graduating students for Pre-MAP and all UW whose first math class was that course, are given in *Table 2*. It should be noted that not all students at UW (or even in Pre-MAP) actually take one of the beginning math classes that we consider here. This can be due to a student majoring in a non-STEM field that does not require any math classes. It can also be due to transfer students passing out of all required math courses with transfer credits.

*Table 2*

| Math Placement | Percent of Total Who | Percent in Course Who |
|---|---|---|

|                                | Started in Course |                       | Graduated in STEM |                     |
|--------------------------------|-------------------|-----------------------|-------------------|---------------------|
|                                | Pre-MAP           | Total STEM Graduates  | Pre-MAP           | Total UW Graduates  |
| Math 098: Algebra              | 1%                | <1%                   | N/A               | 9%                  |
| Math 120: Pre-Calculus         | 21%               | 9%                    | 50%               | 25%                 |
| Math 124: Calculus 1           | 23%               | 21%                   | 75%               | 49%                 |
| Math 125: Calculus 2           | 10%               | 15%                   | 100%              | 62%                 |
| Math 126: Calculus 3           | 22%               | 17%                   | 86%               | 78%                 |
| Math >300: Upper Level Math    | 10%               | 15%                   | 75%               | 90%                 |

*Table 2:* The first column provides a description of each math course. The >300 courses correspond to linear algebra, advanced multivariable calculus, and differential equations. The second and third columns show the fraction of Pre-MAP (second column) students and total (third column) who were initially placed in each math course. These percentages add to less than 100% due to the portion of students that come into UW with transfer credit and do not take a basic math course here. The last two columns give the fraction of Pre-MAP students (fourth column) and total UW graduates (fifth column) that graduate in STEM fields from each math course.

*Table 2* illustrates that Pre-MAP students come from a large variety of math backgrounds. Further, Pre-MAP students that place into lower math courses are more successful at completing a STEM degree than their counterparts (see columns four and five). This could be due in large part because of the mentoring, guidance and support Pre-MAP students receive early in their college careers. By receiving that support, the students are encouraged to persist regardless of initial placement. For Non-Pre-MAP students, the higher math placement increases the likelihood of graduating in a STEM or Physics major. Other than for math 120, the Pre-MAP students show a similar distribution in initial math placement compared with STEM graduates. However, Pre-MAP has a significantly higher fraction of students placing into math 120 compared with STEM graduates.

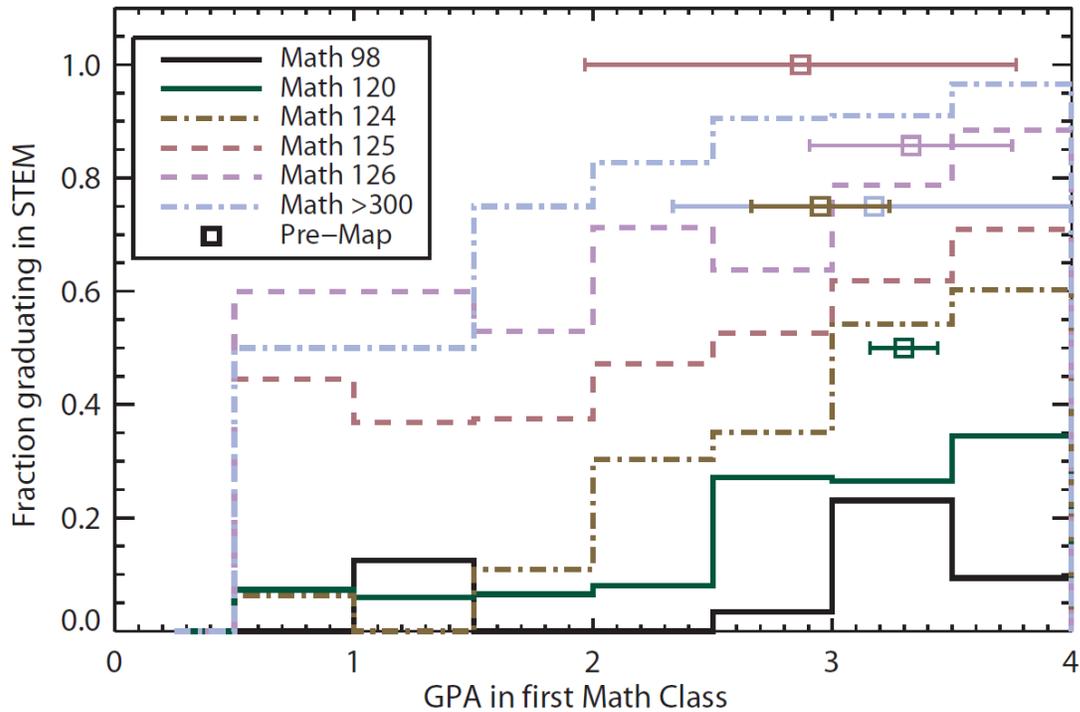

*Figure 3*: Fraction of students graduating in STEM fields as a function of grade in their first math class. Each line denotes a different initial math course. There is a single Pre-MAP point for each math class that is placed at the average grade and STEM graduation fraction for the group of students. The error bars on the data points represent the standard deviation for the gpa of that group of students.

*Table 2* suggests a connection between math placement and likelihood of graduating in a STEM field. We combined data from the student course database with the student graduate database (linked by system key) to investigate final major and GPA compared to first math course. *Figure 3* explores this further by looking at the fraction of students graduating in STEM fields versus the grade in first math classes. *Figure 3* shows that a student's likelihood to major in STEM is greatly affected by their math placement as well as their performance in that math class. Pre-MAP students generally do very well in their first math classes. Further, Pre-MAP students that start in Math 120, 124, 125, and 126, have a higher STEM graduation fraction than other students with similar grades in those classes. Through either a difference in aptitude and interest or through the assistance of the Pre-MAP program, these students are more likely to remain in STEM fields. A notable exception is for students that place in a 300-level math class (307, 308, 309, or 324). These students are all transfer students that entered UW with enough math credit to pass out of the lower level classes. Thus, they represent a very unique population of students. Unfortunately we do not have enough information to understand why Pre-MAP students that place in these classes are less likely than average to pursue STEM majors.

To examine the connection between overall performance with math placement, we compared the average cumulative GPA of graduated students with their math placement finding that Pre-MAP students placing into Pre-Calculus (Math 120) on average have higher cumulative GPAs than all other UW students that also place into Pre-Calculus (*Figure 4*). Pre-MAP students continue to perform about average when placed into other math courses.

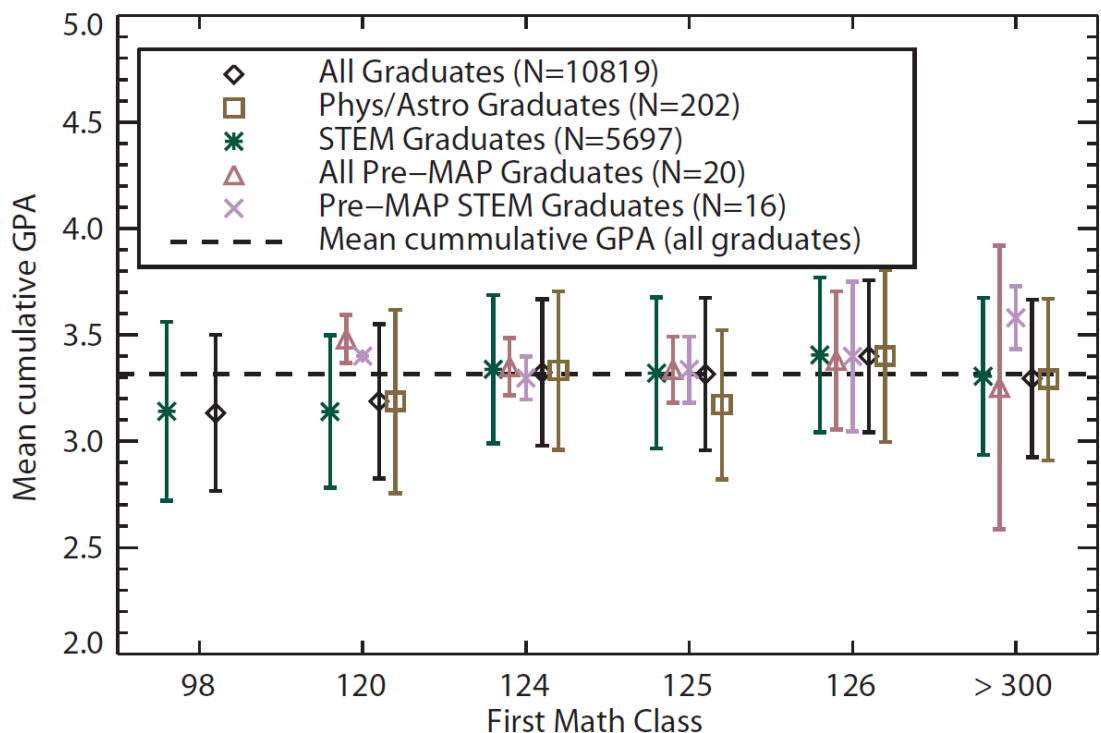

*Figure 4: Average final GPA of graduated students versus their math placement. Pre-MAP students perform similarly to their peers. Pre-MAP students placed in Math 126 and >300 perform better than their peers overall. Error bars represent the standard deviation of GPA for each point.*

When we compare cumulative math GPAs against placement in math classes (*Figure 5*), Pre-MAP STEM majors have math GPAs that are roughly consistent with STEM majors as well as physics/astronomy majors. They tend to perform slightly worse for lower math placements (120-125) and better for higher placements (126, >300), though all within about one standard deviation from the average STEM or physics/astronomy student. We have already shown that math performance for the overall population seems to be an important factor for success in STEM fields. Thus, this can further explain why Pre-MAP students are very likely to major in STEM fields across all math placements.

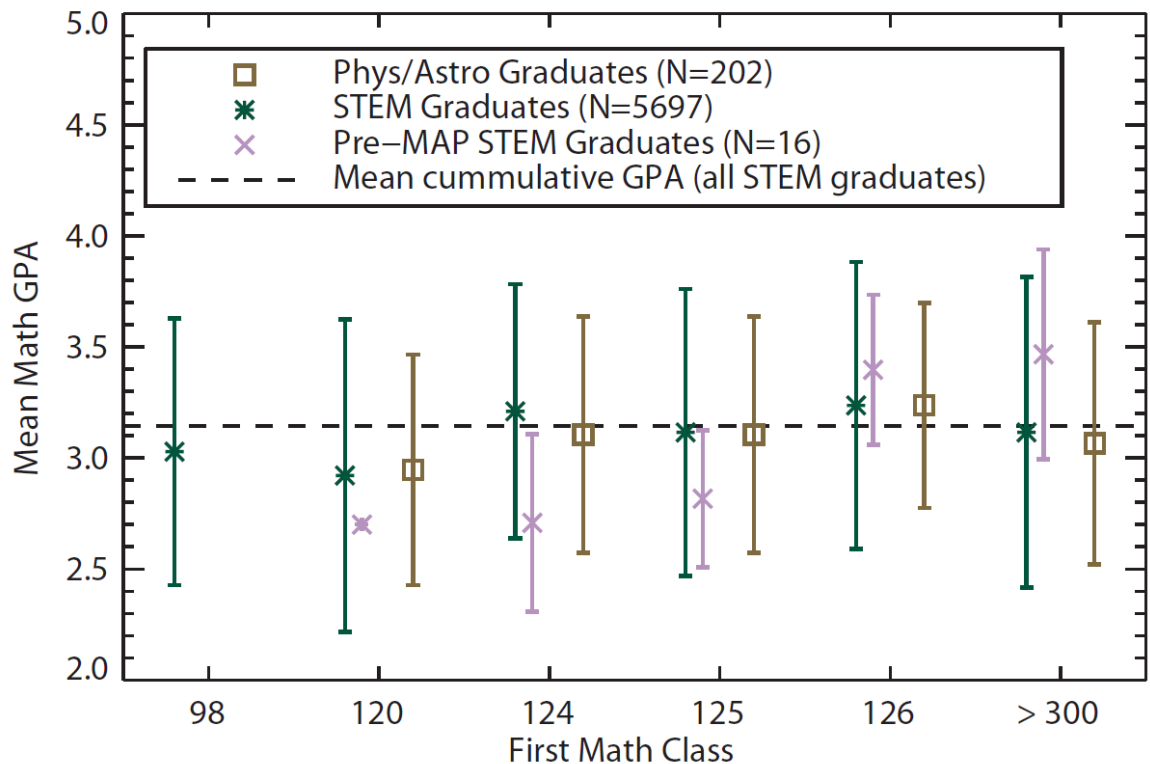

*Figure 5: Average math GPA of graduated students versus their math placement. Note that only Pre-MAP students who graduated in a STEM field are shown. Error bars represent the standard deviation of gpa for each point.*

We also examined the role of high school preparation in the time to degree. We calculated time to degree using entries in the student term major database, which contains an entry for every quarter the student was enrolled. Information on the degree obtained and math course were obtained from the relevant other tables. *Figure 6* shows the average number of quarters it takes students to graduate as a function of first math class. There is a strong trend for all STEM majors, mimicked by the Physics and Astronomy majors, indicating that the lower the level of the first math course, the longer it will take students to graduate. This is due, at least in part, to a delayed start to major course work to which these math courses are a pre- or co-requisite.

The Pre-MAP graduates, both all and STEM, graduate in a similar number of quarters as the overall UW graduates. The dependence of the number of quarters to graduate on first math class is slightly weaker for the Pre-MAP students, but with small numbers we cannot prove that they deviate from UW students as a whole. It is possible that with larger numbers of Pre-MAP students in the future, a trend will become apparent.

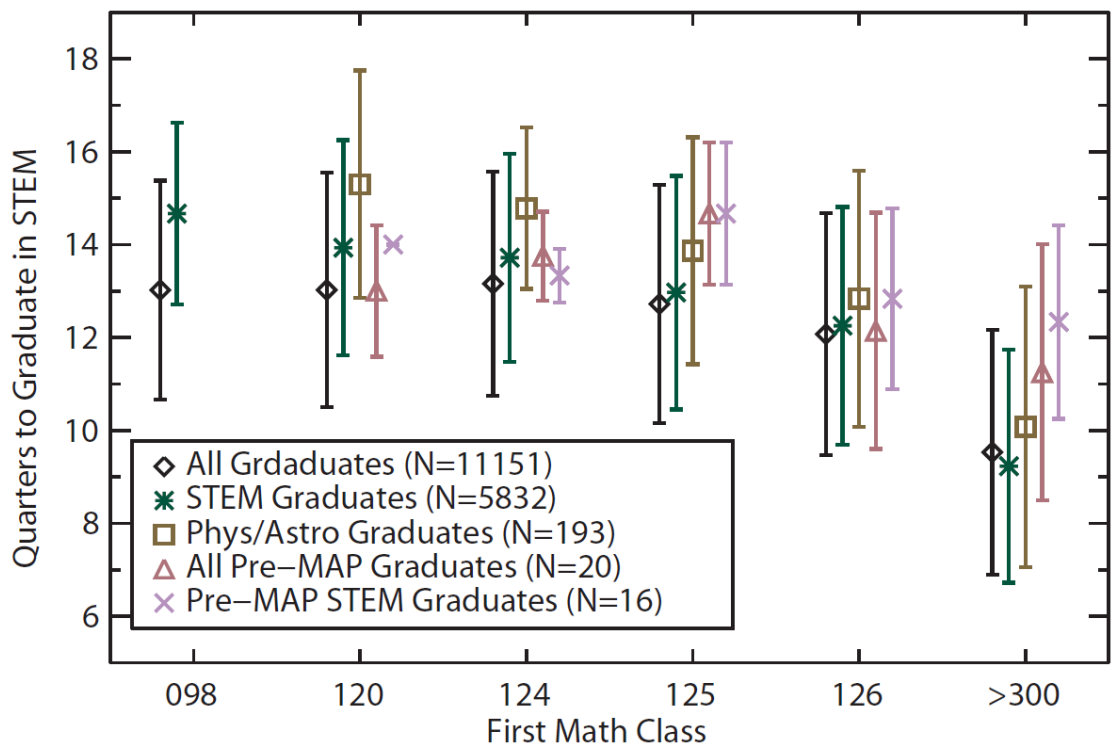

*Figure 6:* The mean number of quarters to graduate for all graduates, STEM graduates, Phys/Astr graduates, and Pre-MAP students as a function of the first math class. The error bars show the standard deviation.

We have shown that math background prior to entering UW, which we infer using the first math class taken by a student, is a crucial factor in determining a student's likelihood to major in a STEM field and how long it will takes them to finish a STEM degree. We find that, while our students place into a wide variety of initial math classes, they are all much more likely to major in STEM fields than their peers. They are also able to perform similarly overall in their classes compared to the overall UW and STEM major populations.

## 5. CONCLUSIONS

This paper has quantitatively compared Pre-MAP students to the broader University of Washington student population, STEM majors in particular, in recruitment, academic performance and graduation. Our guiding questions led us to the following conclusions. Pre-MAP recruits a diverse group of students into the program. In particular, Pre-MAP enrolls students with a variety of math backgrounds who will go on to graduate with STEM degrees at a higher rate than their non-Pre-MAP counterparts. Pre-MAP students also perform well academically compared both with the overall UW population and with STEM students regardless of their initial math placement.

### 5.1 Program Recommendations

*1) Recruit underrepresented students placing into Pre-Calculus:* To make the largest impact as a program, Pre-MAP should focus recruitment on underrepresented students that place into Math 120 – Pre-Calculus AND math 124 – Calculus I.  A shift in recruitment styles in the most recent years may have negatively impacted the diversity of the Pre-MAP student population.  We recommend the Pre-MAP staff reexamine their recruitment strategies to find a solution that is both diverse and not time intensive.  It is this group of diverse Pre-Calculus and Calculus I students that a program like Pre-MAP seems to have the greatest impact on.  At the University of Washington, one suggestion is to strengthen connections with the Office of Minority Affairs and Diversity.

*2) Using math placement as a significant indicator of success:* This research started out investigating broad areas to measure success but the authors quickly realized the important role math preparation plays in a student's retention in STEM.  While math placement should not be used as the sole indicator of success, there does appear to be a strong correlation between them.

3) *Success in math classes should be a priority for Pre-MAP academic advisers.* Math performance is an important indicator of STEM retention so it should be a priority if we want our students to major (and successfully graduate) in STEM fields.  For example, focus study skills development by using math homework and exams.  We would also suggest conducting more frequent academic check-ins for students taking math courses and referring them to math study centers for additional assistance.

**5.2 Future Research**
The authors of this paper came up with many additional questions that were unable to be incorporated into this evaluation paper.  Future evaluation work on Pre-MAP may include the following:

1. Examine the distribution of academically prepared students (compared to Astro, Phys, STEM, University).  For example, incoming math/academic preparation (i.e. high school GPA, math placement, AP scores) compared to cumulative math GPA, counting retakes.
2.  Answer questions related to switching majors such as; Are Pre-MAP students more likely to change their major to something else later in their academic career? Does Pre-MAP retain students in their chosen STEM major?
3. Target future evaluations on 5-10 year post graduation statistics as well as examining specific components of Pre-MAP to determine what might make the largest impact on retention.
5. Another similar, even more in-depth analysis can be done in ~5 years when the number of graduated (and matriculated) students will have greatly increased. Combining quantitative research like this work with focus group and interview data with individual students will also be enormously useful.

**References**
Haggard, D., Garner, S., Cowan, N. 2008 "Pre-MAP: A Case Study in Evaluating Astronomy Diversity Efforts" *Spectrum* Committee on the Status of Minorities in Astronomy, American Astronomical Society, June 2008.

Rosenfield, P., Loebman, S.R., Hilton, E.J., et al. 2010, "Tools for Increasing Undergraduate Diversity in Your Department" Bulletin of the American Astronomical Society, 42, 466.16